\documentclass[runningheads,a4paper]{llncs}

\usepackage{amssymb}
\usepackage{graphicx}
\usepackage{tcolorbox}

\usepackage{url}
\usepackage{xcolor}

\begin{document}

\mainmatter  % start of an individual contribution

% first the title is needed
\title{Blockchain and the Common Good Reimagined\thanks{This paper was prepared for \textit{The Common Good in the Digital Age} conference held in the Vatican City State, between 26-28 September 2019.}}

% a short form should be given in case it is too long for the running head
\titlerunning{Blockchain and the Common Good Reimagined}

% the name(s) of the author(s) follow(s) next
%
% NB: Chinese authors should write their first names(s) in front of
% their surnames. This ensures that the names appear correctly in
% the running heads and the author index.
%
\author{Joshua Ellul and Gordon J. Pace
%\and Ursula Barth
}

\authorrunning{Joshua Ellul and Gordon J. Pace}
% (feature abused for this document to repeat the title also on left hand pages)

% the affiliations are given next; don't give your e-mail address
% unless you accept that it will be published
\institute{Centre for Distributed Ledger Technologies\\
University of Malta\\
Msida, Malta\\
\url{{joshua.ellul, gordon.pace}@um.edu.mt}\\
}

%
% NB: a more complex sample for affiliations and the mapping to the
% corresponding authors can be found in the file "llncs.dem"
% (search for the string "\mainmatter" where a contribution starts).
% "llncs.dem" accompanies the document class "llncs.cls".
%

%\toctitle{Lecture Notes in Computer Science}
%\tocauthor{Authors' Instructions}
\maketitle

%\begin{abstract}
%aaa
%\end{abstract}

\section{Introduction}
Blockchain, Smart Contracts and Distributed Ledger Technology (DLT) are being touted to revolutionise digital services -- through decentralisation. Cryptocurrencies, self-sovereign identities, decentralised certificate registries, and transparent voting systems are but a few applications which promise to empower end users and provide assurances that neither data nor the associated computational logic have been tampered with.

%\gp{Add what is blockchain or rather what it gives}

Satoshi Nakamoto, whoever the individual or group may be, proposed a technology, Blockchain, which can be used to implement a ledger of digital assets ownership in a decentralised manner, and used it to create the first such  cryptocurrency, Bitcoin~\cite{nakamotobitcoin}. Bitcoin's genesis block (the first block mined) included the text \emph{``The Times 03/Jan/2009 Chancellor on brink of second bailout for banks''} --- a clear statement that Bitcoin was intended to challenge the existing banking regime, and may also support claims that Bitcoin emerged as a response to the 2007/8 financial crisis. Decentralised `money' was just the beginning. Shortly afterwards came the advent of smart contracts (as originally conceptualised by Nick Szabo~\cite{szabo1997formalizing} and later implemented on the Ethereum blockchain~\cite{wood2014ethereum}) --- deterministic computer code that can be executed in a distributed and decentralised manner, and which enables new types of trusted, yet decentralised digital processes.

Decentralisation, disintermediation, transparency, verifiability, auditability, openness, inclusion, tamper-proof, immutability are just some of the buzzwords that continue to be swung around in the promotion of the benefits  brought about by Blockchain-based systems to the users. The rhetoric used creates parallels between the features brought about through blockchains and values that many try to uphold, for example honesty, openness, transparency, teamwork and unchanging truth.

From this rhetoric, together with the often heard chant from blockchain evangelists of \emph{``Decentralise Everything!''}, it may be construed that applying Blockchain and DLTs to any process is key to achieve the ultimate common good. However, individual user benefits do not necessarily translate into benefits for the common good --- and as hinted by Aristotle and in the New Testament, it is the noble common good that we should strive towards:

\begin{quote}
``For even if the good is the same for an individual as for a city, that of the city is obviously a greater and more complete thing to obtain and preserve.'' -- Aristotle \cite{aristotle}
\end{quote}

\begin{quote}
``\textsuperscript{3} Do nothing out of selfish ambition or vain conceit. Rather, in humility value others above yourselves, \textsuperscript{4} not looking to your own interests but each of you to the interests of the others.'' -- Philippians 2:3--4 (NIV)
\end{quote}

%{\color{red} add citation to verse?}

In this paper a number of blockchain applications aimed at supporting initiatives for common good are highlighted. This is followed by a discussion on technology de/centralisation and a thought experiment used to raise questions regarding the use of decentralised technology in terms of social implications. We close the paper with some parting thoughts, highlighting the more important questions and challenges we leave pending.

\section{Common Good Blockchain Use Cases}
\label{s:examples}

Three typical types of use cases will now be presented where blockchain systems can be used for common good including: providing a means of providing banking for the unbanked; implementing mechanisms to assure fair and ethical processes; and community empowerment.

\subsection{Financial Inclusion}
According to the Global Findex Database issued in 2017 by the World Bank \cite{demirguc2018global} 1.7 billion adults worldwide (corresponding to 31\% of adults) are unbanked. Having no access to a bank account means they may not be able to get access to potential important services that may require bank statements when signing up. The main factors for the unbanked can be classified as: (i) logistical or financial reasons including limited or no physical access to financial institutions, not having enough funds to open or operate an account, costs of operating an account are too expensive, and lack of identification; or (ii) social reasons including reliance on family members who have accounts, religious reasons, and lack of trust in financial institutions \cite{lichtfous}. 

It is also reported, however, that of these unbanked adults, two-thirds own a mobile phone. The use of cryptocurrencies and blockchain through such devices can overcome the logistical, financial and social hurdles listed. Blockchain and DLTs deal precisely with disintermediation and the removal of central points of trust previously required, and may thus have the desired characteristics for individuals that lack trust in financial institutions to be more open to using cryptocurrencies. Depending upon the underlying religious belief that is stopping someone from using a bank account, cryptocurrencies may or may not provide a solution for them to gain access to bank-like features using cryptocurrencies. There is however little that cryptocurrencies can do for individuals who are reliant on family members, though access to blockchain based services may become more accessible.

Although cryptocurrencies and blockchain-based solutions could provide the technical mechanism to overcome these issues, depending upon the jurisdiction they may only provide a partial or temporary solution. As cryptocurrencies (and other money-like assets) may be subject to the law of the land which may outright ban them, regulate them, or simply say nothing at all about them. Having a regulatory framework in place which addresses such assets based on their particular features is obviously an ideal situation as it provides legal certainty to end users, enabling them to understand rules and potential risks. Outright banning cryptocurrencies or blockchain-based systems that could enable financial inclusion in a jurisdiction is a show-stopper, at least for those individuals that want to abide by the law. Whilst jurisdictions that do not provide crypto-regulation do allow for the unbanked to gain financial inclusion, but with the risk that this may be only temporary until the jurisdiction puts such regulation into place. This creates an uneasy environment whereby cryptocurrency owners may fear that their assets would be confiscated or devalued.

\subsection{Assuring Fair and Ethical Supply Chains}
Human rights, fair trade, environmental friendliness and other ethical terms and initiatives have been making headlines and surrounded our everyday lives for a while now --- but are we there yet?

In a 2018 Nature Sustainability paper Nkulu et al.~\cite{nkulu2018sustainability} reveal how those working and living in areas surrounding cobalt\footnote{Cobalt is a metal used for lithium-ion batteries.} mining areas had much higher levels of cobalt in their blood and urine, of which \emph{``were most pronounced for children''}, in some of whom evidence was also found of DNA damage.
%https://www.ibm.com/blogs/industries/blockchain-cobalt/

A 2017 report~\cite{international2017global} highlighted that 152 million children are victims of child labor, out of which 73 million work in hazardous conditions.

%{\color{red}blood diamonds, slavery, funding of oppressive regimes, coffee farmer poverty}

%https://www.forbes.com/sites/bernardmarr/2018/03/14/how-blockchain-could-end-the-trade-in-blood-diamonds-an-incredible-use-case-everyone-should-read/#56f5a4aa387d

When we buy products, typically, we need to trust that fair and ethical considerations and processes were used amongst the various different stakeholders involved in the particular supply chain. Over the past decades authorities, standards bodies, non-governmental organisations and other institutions worked at implementing procedures and standards which would provide more assurances in respect to such fair and ethical processes. However, often such processes include spot-checks, and sampling of products and services which still leave room for foul play --- since only a sample would be checked suppliers could mix in both licensed and unlicensed products together.

%{\color{red} add point on can only assure trust in the decentralised digital world -- as soon as we go to the real world, it's lost.}
%{\color{red} garbage in-garbage out}

A blockchain based-supply chain solution could enable each stakeholder to input and/or verify the work they have done and materials used. This would provide for a tamper-proof source to end-product trace of each process and material used in a product. That being said, such a solution would be dependent on each stakeholder entering true information. Such audit processes described above could put in place to check whether stakeholders are indeed entering the correct information to help minimise such activity.

Internet of Things (IoT) and other connected sensor devices can be used to automatically detect real world physical supply chain events and store them into the associated blockchain. However, it is important to note that such IoT devices could also be subject to submitting erroneous or missing data whether intentionally or not. Indeed, there are ways of addressing this: using secure and tamper-proof devices, increasing numbers of devices for redundancy, and allowing for different stakeholders to use their own devices to guarantee their respective interests.

%food safety 
%U.S retail giant, Wall-mart, Tsinghua University and IBM in Beijing are testing the blockchain technology to tackle food safety issue in China (Kharif 2016). Thanks to the use of the blockchain technology, consumers can trace back where the pork came from within China, thus certifying and ensuring its quality. The blockchain technology becomes more useful when it combines with the recent IoT technologies such as RFID sensors or QR codes. Without extra manual work to upload information on the blockchain database, the process of certifying the food quality can be automated

\subsection{Community Empowerment}
Over time corruption is either becoming more prevalent, more visible to the public eye (perhaps due to social media), or even being announced before sufficient (or any) evidence is available, as fake news become rampant. 

A 2014 OECD report on foreign bribery states that 57\% of bribery cases were to secure public procurement contracts \cite{oecd2014oecd}.

Rigging in voting from board motions, to local community affairs, to national elections is a topic that keeps resurfacing~\cite{jimenez2017testing}.

Computerised systems could help in such situations to both provide for more efficient processes as well as to put mechanisms in place to minimise potential manipulation as much as possible. However, as long as the systems are operated, owned or maintained by centralised entities, room is left for interference. Often independent audits of systems are undertaken to provide assurances, however ultimately because of their centralised nature such assurances cannot ultimately be provided --- and even if a system could be shown to be tamper-proof in such a case, the public may still doubt it due to the inherent centralised control of power.

Smart contracts can provide a solution to the above scenarios, by implementing a tamper-proof, verifiable, auditable, publicly open system that can enforce the required computational logic and assure that neither the logic nor data can be tampered with. Indeed, for many types of voting systems it would not be ideal to have individual users' votes disclosed. Using techniques such as zero-knowledge proofs~\cite{goldwasser1989knowledge} it is possible to both keep secret individuals' votes whilst assuring that the outcome cannot be tampered with.

\section{Technology De/centralisation}
\label{s:decentralisation}
From the use cases described above one can note how decentralisation and with it tamper-proof, immutable, verifiable ledgers and systems can allow for opportunities to work towards the common good. 

Decentralisation of applications, that is the complete opening and disclosure of data and logic --- does not leave room for questions to be raised with respect to trustworthiness of the system, due to it being completely transparent. That is of course, if an application is completely decentralised --- and decentralisation is often believed and described to be a binary option, in that either the whole system is decentralised or centralised, both from a technology implementation point of view and from a social governance point-of-view. The truth though, is that unless a decentralised application (dApp) operates completely within a decentralised environment, then it cannot be fully decentralised technology-wise. Furthermore, even then, other points of centralisation may exist --- for instance in the logic, which may require centralised confirmation throughout its operation (for example from an administrator).

Identifying various points of de/centralisation and their impact in achieving the common good goals of the particular application is crucial, since certain centralised points could ultimately defy the whole scope of the application being used for common good and result in a waste of funds, be it public or private funds, which in and of itself would not be in the interest of the common good. %A description of the various relevant technology layers will follow, so as to be able to explore various levels of de/centralisation within dApps, which will be followed by a discussion on levels of decentralisation within social governance.

%\subsection{Decentralisation Considerations}
%Figure~\ref{fig:dltlayers} depicts an abstract view of the different software layers which make up a dApp\footnote{The figure reproduces the software layers from the Malta Digital Innovation Authority's technology stack as defined in the Technology Stack Nomenclature guidelines available from \url{https://mdia.gov.mt/guidelines/}}. 

%\begin{figure}
%    \centering
%    \includegraphics[scale=0.55]{images/VaticanTechStack.png}
%    \caption{dApp Software Layers.}
%    \label{fig:dltlayers}
%\end{figure}

%At the very bottom, the DLT Implementation Layer represents the various individual computers or nodes, that are working together to keep the distributed and decentralised ledger in check -- in the case of a system like Ethereum or Bitcoin this layer represents the Blockchain. This may include nodes acting as \textit{miners} or \textit{verifiers} or any other particular functionaries which the particular DLT requires to assure consistency and validity of the shared ledger. This layer can be seen to be the logic that allows for distributed and decentralised nodes to operate together to provide a single shared abstract ledger or computational machine to external users and/or systems. 

A blockchain or DLT represents various individual computers or nodes, that are working together to keep the distributed and decentralised ledger in check. This may include nodes acting as \textit{miners} or \textit{verifiers} or any other particular functionaries which the particular DLT requires to assure consistency and validity of the shared ledger. This layer can be seen to be the logic that allows for distributed and decentralised nodes to operate together to provide a single shared abstract ledger or computational machine to external users and/or systems. 

Blockchain systems rely on majority of the network of nodes to operate correctly to achieve consensus on the canonical ledger --- though work shows that a 51\% majority may often not be enough~\cite{eyal2018majority}. When we speak about decentralisation at the DLT or blockchain level what is really implied is that anyone who wants to take part in the network can do so. In reality though, what level of decentralisation actually emerges? In aim of sharing mining rewards, individual miners often pool together to have a higher collective chance of sharing the reward. However, once mining pools achieve substantial combined computational power, one could argue that there is heavy reliance on the particular mining pool for correct operation of the blockchain -- that is centralisation of power. This view applies not only at the grouping of a mining pool, however one could also view this issue at the grouping level of a jurisdiction --- which may have its own beliefs, laws, social norms, etc. In fact many have raised the question whether blockchain systems like Bitcoin provide decentralisation \cite{gervais2014bitcoin} when not only considering technological decentralisation but considering social decentralisation aspects as well. The participating miners in a blockchain work towards achieving the \textit{common good of the participating miners} which does not necessarily translate into the \textit{common good of the users of the system} or, even more importantly, to that of \textit{the common good}. 

A smart contract can be considered to be nothing more than a computer program (with particular attributes) -- an app or application that executes on top of a blockchain. The smart contract code executes on top of the decentralised DLT (or Blockchain) however whether a smart contract provides for decentralisation or not is dependent on its code. The DLT provides an abstract `One World Computer' (the abstract view of considering that all DLT nodes working together are providing a single computer which executes programs, or smart contracts), and the smart contracts are the computational logic that execute on top of this `one world computer' which nobody can tamper with (they will do exactly what the code says they will do). 

When writing (or using) smart contracts one needs to consider what benefits are really being provided in terms of de/centralisation. Whilst the code is tamper-proof and will execute exactly what it is written to do so, does the particular smart contract enable for democratisation of power for the particular service? Or is power still held by a central authority who is the smart contract's super-user? Really, this is lesser a question of technology implementation, and more a question of social governance. In aim of working towards the common good, what powers and decisions should be democratised or made by the community or users and what powers and decisions should be kept at a central authority?

\section{Enforced Social Contracts: A Thought Experiment}
\label{s:thought-experiment}
Agreements between parties, whether legal or otherwise, speak about the ideal state of affairs. As opposed to classical logic, in which statements refer to the actual state of affairs, agreements speak about the way things should be, identifying rights and obligations on the parties involved. An obligation to deliver an item before a deadline is no guarantee that the act will actually take place, and many agreements include reparation clauses addressing when one of the parties fails to carry out their obligations. In case of a breach not catered for by a reparation within the agreement itself, the fallback is other regulatory structures be it formal (the law) or otherwise (social pressure).

Smart contracts were proposed by Szabo as a step forward in that the contract is self-executing or enforcing, such that they \emph{``make breach of contract expensive (if desired, sometimes prohibitively so) for the breacher''}~\cite{szabo1997formalizing}. One can argue that smart contracts have existed millennia before Szabo's proposal in 1997. If the parties involved in an agreement identify a trusted third party, then execution of the contract can be entrusted to them --- escrows, for instance, work precisely in this manner. One could argue, however, that the third party is just another implicit party in the agreement, and nothing stops them from breaching the confidence entrusted to them by the parties. Szabo's vision was to do away with such trusted parties, a vision which was only realised with the rise of the second generation of blockchain solutions~\cite{wood2014ethereum}, allowing for the trusted execution of logic (code) without the need for such trusted third parties. 

When it comes to community-level regulation --- social contracts, whether through formal law or through social norms, the situation is similar to that of legal contracts. Nothing stops individuals from following behaviour going against the social contract. Indeed, Socrates argues at length on why he \emph{should} act as prescribed by the social contract in Plato's \emph{Crito}~\cite{plato}. The dialogue hinges on the fact that one has a \emph{choice} as to whether or not to follow a just law, based on, or despite of, the state's (or society's) demand for loyalty to its rules in the form of an underlying social contract. One can argue that Socrates' view is that the social contract one partakes in by being a member of a community softly enforces that community's underlying legislation. We use the term \emph{soft enforcement} to highlight the fact that, just like the underlying laws, nothing stops individuals from going against the social contract. This goal of this social pact has been argued by Rousseau~\cite{rousseau} and later writers, to be that of acting in the common good of that society.

The natural question that arises is on the use of technology to enforce legislation and the underlying social contract in order to maximise the common good. Smart contract technology can certainly act as the required tool to enable such \emph{hard enforcement}. In this context it is worth mentioning the DAO (a Decentralised Autonomous Organisation)~\cite{dao}, which can be seen as having been a social experiment in setting up an organisation in which decision taking is decentralised according to fixed rules in a smart contract running on Ethereum, ensuring that the behaviour of the organisation follows the will of its members. Although limited in scope, it can be seen as an exploration of how social rules on decision taking can be enforced without a centralised authority\footnote{Perhaps it is worth noting here that the DAO failed because of technology malfunction --- a bug in the code of the underlying smart contracts. This brings to the fore one of the risks of technology-based enforcement of societal rules, but we will briefly come back to this in the final section of this paper.}, and indications are that the scope of such organisations could be widened using similar technologies.

The question, which remains unanswered though, is whether such hard enforcement is a desirable means of nudging or pushing the community towards the common good. We do not claim to have an answer to this. However, we will present a thought experiment in order to highlight why the answer to this question is not as straightforward as one may initially hazard to think. Consider the following setting for a thought experiment.

\small\begin{tcolorbox}
\noindent\textbf{Setting:} \emph{Consider a fictitious land in which everyone lives in town $A$ and works in town $B$. There are two routes between the two towns (i) a short route which typically requires a 10 minute drive, but becomes clogged when more than $N$ cars take it at the same time, resulting in a 20 minute trip; and (ii) a longer route which always takes 25 minutes no matter how many cars take it. We will assume that in any given day everyone leaves work at the same time, and only $n$ persons (with $n < N$) need to arrive to work within 10 minutes of their departure (with the persons who need to arrive early changing from one day to the next).}
\end{tcolorbox}\normalsize

Let us start our thought experiment by resolving the use of the limited resource by allowing individual choice. At an individual level and that of the community, it is desirable that since the fast route suffices for those in a hurry, then its use is rationed and no one will ever arrive late to work. 

\small\begin{tcolorbox}[standard jigsaw,opacityback=0]
\noindent{\textbf{Resolution 0:}} If everyone takes the fast route only when they really need it, then the system works, with everyone arriving to work on time. But an individual may reason that if they are the only ones taking the faster route, then that route will still not be clogged. But the tragedy of the commons \cite{hardin1968tragedy} hits, with everyone reasoning in this manner, and resulting in clogging of the fast route. Everyone ends up spending 20 minutes in traffic, with all those in a hurry arriving late. Furthermore, no one has any incentive to switch to the slower route on the following day, since that will anyway take longer.
\end{tcolorbox}\normalsize

Since the ideal behaviour is unlikely to emerge, we can look at ways of enforcing ideal behaviour in order to ensure that the behaviour of individuals is to the benefit of the community as a whole. We will start with the use of a smart contract style approach imposed on the use on all cars.  

\small\begin{tcolorbox}[standard jigsaw,opacityback=0]
\noindent{\textbf{Resolution 1:}} Now imagine a social contract encoded as a smart contract controlling everyone's self-driving car (with no manual override), deciding which route to take in a common good, yet decentralised manner: All those in a hurry together with $N-n$ of the rest (randomly chosen) taking the fast route. The result is that all those in a hurry (and some others) will be arriving to work in just 10 minutes. 
\end{tcolorbox}\normalsize

\emph{Resolution 1} guarantees not only that all individuals will arrive on time to work, but also that the use of the fast route is maximised and its use remaining capacity is distributed in a just and fair manner. From a utilitarian perspective, this is clearly an ideal route usage policy --- by hard enforcing the the social contract that those not in a hurry should allow those who are to use the fast route, the community as a whole benefits, and indicates a gain over the previous soft enforcement. There are, however, other means of enforcing the social contract.

\small\begin{tcolorbox}[standard jigsaw,opacityback=0]
\noindent{\textbf{Resolution 2:}} Now imagine if this fictitious land were a dictatorship, and instead of the social smart contract, a device is to be connected to everyone's car such that whenever the fast route is delayed, all those not in a hurry and who have taken it will be thrown in jail for a week. We will assume that one cannot know whether the fast route is delayed unless one has already commited onself to it.
\end{tcolorbox}\normalsize

\emph{Resolution 2} is a soft enforcement implementation, in that individuals still have a choice, albeit having a harsh disincentive to act selfishly. The underlying threat of a week in jail imposed on the population in order to achieve the desirable outcome is, however, not an ideal means of achieving the result. Many would consider this to be a disproportionate means to an end. However, the fact remains that this is a soft enforcement, and the hard enforcement of the smart contract is certainly an even more draconian means of achieving the result. If one were to consider \emph{resolution 2} to be an undesirable, perhaps even unethical solution, \emph{resolution 1} should be perceived to be even less acceptable. And yet, paradoxically, many would be ready to accept \emph{resolution 1} though not \emph{2}. One aspect common to both these resolutions is that the compulsory participation in the system. The question is thus whether the smart social contract solution would be more ethical to adopt if participation is optional. 

\small\begin{tcolorbox}[standard jigsaw,opacityback=0]
\noindent{\textbf{Resolution 3:}} Finally, imagine a situation in which participation in the social smart contract is on a voluntary basis, although once joined, leaving is no longer an option, and with the enforced agreement taking only into consideration individuals who have already joined when distributing cars to routes. Consider an individual who has not yet signed up. If more than $N$ persons have already signed up, she knows that she will never arrive to work on time when in a hurry (since the social smart contract will be assigning $N$ persons to the fast route). Clearly, this means that she has an incentive to join. On the other, if there are less than $N$ participants, she knows that by joining she is signalling to everyone that she has no choice but to take the fast route (since the contract will assign the first $N$ to that route), meaning that others will make a choice based on this knowledge, and possibly try to optimise usage altruistically or go for the fast route egoistically. Either way, she is better off, since the behaviour of others will not be changing for the better. In other words, it is always advantageous to join. 
\end{tcolorbox}\normalsize

Participating in the smart social contract of \emph{resolution 3} is voluntary, and thus not an imposed social norm. Perhaps this is a way of addressing the draconian nature of compulsory participation. And yet, rationally speaking, everyone's best choice is to join, thus resulting in effectively being equivalent to \emph{resolution 1} (under the assumption that all potential participants are rational). Does a choice with an option which is rationally better than another really  constitute choice, or is it simply an imposition of that option? 

In this section we have used a variant of a thought experiment typically used to illustrate the phenomenon of the tragedy of the commons, to show how technology-driven enforcement of social contracts raises various philosophical questions and challenges which need to be answered. The means of achieving the common good are not always justified, and technological possibility is not a sufficient cause for adoption.  

%\section{Full Decentralisation and Democratisation}
%Imagine a World, where 

%\section{Opportunities and Challenges for\\Blockchain for the Common Good}
%bugs in code... requirement for assurances

%{\color{red} Technology challenges, are we there yet?  %Trust/ssl example, QR codes etc, usage, etc etc}\\
%{\color{red} Objective/Subjective-ness - what r ppl agreeing to?}\\
%{\color{red} -	Blockchain/DLTs can hide individuals – how can we lift the veil where required?}\\
%{\color{red} Blockchain, Energy Costs, Other DLTs, etc}\\
%{\color{red} Should we even decentralise/democratise certain processes?}\\

\section{Parting Observations and Thoughts}
Technology is frequently argued to be an enabler of or a means for achieving a better world, somehow improving on the common good. However, nothing is unidimensional, and in the end it is always a measure of the opportunities and threats brought about by technology. In this paper we have attempted to make observations and raise questions regarding the role blockchain can play towards the common good. Blockchain-technology was never value-neutral. The original paper presenting the technology vying away from traditional means (not an industrial whitepaper or academic publication, by an anonymous author), and is more of a (technical) manifesto with a sociopolitico-technical war-cry of \emph{``Down with points of trust''}. Even the genesis block of the first blockchain implementation contains a politically-laden message embedded inside it. Clearly, the opportunities brought about by the technology are great, as the real-world use cases we illustrated in Section~\ref{s:examples} amply show. These opportunities largely revolve around decentralising trust, decentralising control, although these in themselves are broad terms with different interpretations or points of relevance as we have discussed in Section~\ref{s:decentralisation}. 

One aspect of decentralised control is that of decentralised logic, which has been used to enforce protocols of interaction between participants in the form of smart contracts. It remains to be seen whether pushing such contract enforcement to the level of social contracts is a good idea. In Section~\ref{s:thought-experiment} we present a thought experiment raising questions regarding the ethics of doing so but shy away from providing answers. 

Many question, however, whether when it is a technology that does away with human or social trust requirements, whether we are simply pushing the trust onto the technology itself. Is it simply that we are now trusting the technology? And should we do so? As mentioned in Section~\ref{s:thought-experiment}, the DAO failed due to technology malfunction --- a software bug under which the whole `social' experiment collapsed. The solution used to respond to that crash was to push the trust back outside of the technology --- asking miners and validators to forget all that went wrong and rewind time back to the Garden of Eden before the fruit of knowledge was bitten into through a blockchain fork.

%\subsubsection*{Acknowledgments.} IVLP....

\bibliographystyle{apalike}
\bibliography{refs}

%\begin{thebibliography}{4}
%\end{thebibliography}

%\section*{Appendix: Springer-Author Discount}
%Traceability
%Accountability
%Appendix

\end{document}